\documentclass[review,12pt]{elsarticle}

\usepackage{amssymb}
\usepackage{amsthm,amsmath}
\usepackage{accents}
\usepackage{graphicx}
\usepackage{epstopdf}
\newtheorem{theorem}{Theorem}

\newtheorem{Definition}{Definition}

\newtheorem{remark}{Remark}

\allowdisplaybreaks

\begin{document}

\begin{frontmatter}

\title{Finite time Convergence of Pinning Synchronization with Linear and Nonlinear Controllers\tnoteref{tt}}\tnotetext[tt]
{This work is jointly supported by the National Key R$\&$D Program of China (No.
2018AAA010030), National Natural Sciences Foundation of China under Grant (No.
61673119 and 61673298), STCSM (No. 19JC1420101), Shanghai Municipal Science and
Technology Major Project under Grant 2018SHZDZX01 and ZJLab, the Key Project of
Shanghai Science and Techonology under Grant 16JC1420402.}

\author[lwl-1,lwl-2,lwl-3,lwl-4]{Wenlian Lu}
\ead{wenlian@fudan.edu.cn}
\author[lxw-1]{Xiwei Liu}
\ead{xwliu@tongji.edu.cn}
\author[ctp-1,lwl-1]{Tianping Chen\corref{tpchen}}
\ead{tchen@fudan.edu.cn}

\address[lwl-1]{School of Mathematical Sciences, Fudan University, Shanghai 200433, China}
\address[lwl-2]{Institute of Science and Technology for Brain-Inspired Intelligence, Fudan University, Shanghai 200433, China} \address[lwl-3]{Shanghai Key Laboratory for Contemporary Applied Mathematics and Laboratory of Mathematics for Nonlinear Science, Fudan University, Shanghai 200433, China}
\address[lwl-4]{Shanghai Center for Mathematical Sciences, Fudan University, Shanghai 200433, China}
\address[lxw-1]{Department of Computer Science and Technology, Tongji University, Shanghai 201804, China}

\cortext[tpchen]{Corresponding author. E-mail address: tchen@fudan.edu.cn}

\begin{abstract}
In this paper, we propose several models, which can realize synchronization of complex networks in finite time effectively. The results apply to heterogeneous dynamic networks, too. The mechanism of finite time convergence is revealed. Different from many models, which assume the coupling matrix being symmetric (or the connecting graph is undirected), here, the coupling matrix is asymmetric (or the connecting graph is directed).
\end{abstract}


\end{frontmatter}


\section{Introduction}
Synchronization of complex networks has been a hot topic in recent decades. The model in the synchronization literature can be described as
\begin{align}
\dot{x}_i(t)=f(x_i(t))+c\sum_{j=1}^{m} a_{ij}\Gamma(x_j(t)-x_i(t))
\end{align}
where $x_{i}(t)=(x_{i}^{1}(t),\cdots,x_{i}^{n}(t))^{T}\in R^{n}$, $i=1,\cdots.m$.

Under previous coupling protocols, various sufficient criteria for
asymptotic or exponential synchronization can be obtained. 

Pick $\varepsilon>0$, in [1], the following coupled network
with a single controller
\begin{eqnarray}\label{pina}
\left\{\begin{array}{cc}\frac{dx_1(t)}{dt}&=f(x_1(t),t)
+c\sum\limits_{j=1}^ma_{1j}\Gamma x_j(t)\\
&-c\varepsilon\Gamma(x_{1}(t)-s(t)),\\
\frac{dx_i(t)}{dt}&=f(x_i(t),t)+c\sum\limits_{j=1}^ma_{ij}\Gamma x_j(t),\\
&i=2,\cdots,m \end{array}\right.
\end{eqnarray}
was proposed. It can pin the complex dynamical network $(1)$ to $s(t)$ with $\dot{s}(t)=f(s(t))$ exponentially, if $c$ is chosen suitably.

Denote $\delta x_i(t)=x_i(t)-s(t)$, $\delta f(x_i(t))=f(x_i(t),t)-f(s(t))$, then the system $(1)$ can be rewritten as:
\begin{eqnarray}
\frac{d\delta
x_i(t)}{dt}=\delta f(x_i(t))+c\sum\limits_{j=1}^m{a}_{ij}\Gamma\delta
x_j(t)
\end{eqnarray}
and the network with a single controller $(2)$ is written as
\begin{eqnarray}
\frac{d\delta
x_i(t)}{dt}=\delta f(x_i(t))+c\sum\limits_{j=1}^m\tilde{a}_{ij}\Gamma\delta
x_j(t),~ i=1,\cdots,m
\end{eqnarray}
where $\tilde{a}_{11}=a_{11}-\varepsilon$, $\varepsilon>0$ and
$\tilde{a}_{ij}=a_{ij}$ otherwise.

The topic finite time convergence, finite time synchronization, consensus have attracted attentions of many researchers. Detail introduction and references have been given in [3]-[9], and references therein.

In [9], by using the theory of Filippov, Cohen-Grossberg
neural networks with monotone increasing discontinuous activation functions was discussed. It is revealed that if the equilibrium $x^{*}$ lies in the discontinuity of the activation functions, then finite time
convergence can be ensured (see [9], Th. 8). Cort$\acute{e}$s [9] considered $\dot{x}=-sgn(grad(f)(x))$ and proved the finite time stability and applied it on the network consensus problem. It is clear that the equilibrium $x^{*}$
satisfying $grad(f(x^{*}))=0$ lies in the discontinuity of the function $sign(x)$.
In this paper, based on the idea, we propose simple finite time synchronization models and prove that under mild conditions, they can reach finite time synchronization. 

\section{Some basic concepts and background}

\begin{Definition} A coupling matrix $A=(a_{ij})_{i,j=1}^{m}$ is defined as a Metzler matrix with zero row sum, i.e. satisfying
$a_{ij}\ge 0$, if $i\ne j$ and $a_{ii}=-\sum_{j\ne i}a_{ij}$.
\end{Definition}
\begin{Definition}  A coupling matrix with a single pinning control on the first node $\tilde{A}=(\tilde{a}_{ij})_{i,j=1}^{m}$ is defined as follows
$\tilde{a}_{11}=a_{11}-\varepsilon$, $\varepsilon>0$ and
$\tilde{a}_{ij}=a_{ij}$ otherwise.
\end{Definition}
Let $[\xi_{1},\cdots,\xi_{m}]^{T}$ be the left eigenvalue of the
matrix $A=(a_{ij})_{i,j=1}^{N}$. It is well known that (see [2]) if $A$ is irreducible and
$Rank(A)=m-1$, then, all $\xi_{i}>0$, $i=1,\cdots,m$.

Define $\Xi=diag[\xi_{1},\cdots,\xi_{m}]$. The symmetric part of
$\frac{1}{2}(\Xi\tilde{A}+\tilde{A}^{T}\Xi)$ is negative definite with eigenvalues $0>\mu_1\ge\mu_{2}\ge\cdots\ge\mu_m$ (see [1]).

By M-matrix theory, there are constants $\theta_{i}>0$, $i=1,\cdots,m$,  such that 
\begin{align}\label{M1}
\sum_{i=1}^{m}\theta_{i}\tilde{a}_{ij}<-\theta,~~\theta>0
\end{align}
for all $j=1,\cdots,m$.

Without loss of generality, in the following, we always assume that $\sum_{i=1}^{m}\theta_{i}=1$. 

\begin{Definition} (QUAD function) A function $f(x,t)$ is called $f\in QUAD(P,\Phi,\eta)$, if
\begin{equation}\label{QUAD}
(x-y)^{\top}P(f(x,t)-f(y,t)-\Phi x+\Phi y)
\leq -\eta (x-y)^{\top}(x-y)
\end{equation}
where $P$ is a positive definite matrix, $\Phi$ is a matrix, $\eta>0$ is a constant.
\end{Definition}
\section{Finite time synchronization}

In this section, we propose several synchronization models and prove useful theorems.

\begin{theorem} Suppose that $\Gamma=diag[\gamma_{1},\cdots,\gamma_{m}]$ is a positive diagonal matrix. $||f(x)-f(y)||_{1}\le L||x-y||_{1}$.
Then, the controlled system
\begin{align}\label{pinf1}
\dot{x}_{i}(t)-\dot{s}(t)&=f(x_{i}(t))-f(s(t))\nonumber\\
&+c_{0}\sum\limits_{j=1}^{m}\tilde{a}_{ij}\Gamma (x_{j}(t)-s(t))-\frac{ x_{i}(t)-s(t)}{||x_{i}(t)-s(t)||_{1}}
\end{align}
can synchronize to $s(t)$ in finite time if $c_{0}\theta>L$. Here, $||x_{i}(t)-s(t)||_{1}=\sum_{k=1}^{n}|x_{i}^{k}(t)-s^{k}(t)|$. 
\end{theorem}
{\bf Proof}~
Notice $\delta{x}(t)=[\delta{x}_{1}(t),\cdots,\delta{x}_{m}(t)]^{T}$, $\delta{x}_{i}(t)=[\delta{x}_{i}^{1}(t),\cdots,\delta{x}_{i}^{n}(t)]$, $i=1,\cdots,m$. $||\delta{x}_{i}(t)||_{1}=\sum\limits_{j=1}^{n}|\delta{x}_{i}^{j}(t)|,$
$||\delta{x}(t)||_{1}=\sum\limits_{i=1}^{m}||\delta{x}_{i}(t)||_{1}$. $\delta f(x_{i}(t))=f(x_{i}(t))-f(s(t))$.

With the constants $\theta_{i}>0$ satisfying inequality (5), define a norm
\begin{eqnarray*}
||\delta{x}(t)||_{\{\theta,1\}}=\sum_{i=1}^{m}\theta_{i}
||\delta{x}_{i}(t)||_{1}
\end{eqnarray*}
where $\theta$ and $\theta_{i}$ are defined in (5).

Differentiating, by $||f(x)-f(y)||_{1}\le L||x-y||_{1}$, we have
\begin{align*}
&\frac{d
||\delta{x}(t)||_{\{\theta,1\}}}{dt}=\sum\limits_{i=1}^{m}
\theta_{i}\frac{d
||\delta{x_{i}}(t)||_{1}}{dt}=\sum\limits_{i=1}^{m}\sum\limits_{k=1}^{n}
\theta_{i}\frac{d
|\delta{x_{i}^{k}}(t)|}{dt}\\
&=\sum\limits_{i=1}^{m}\sum\limits_{k=1}^{n}
\theta_{i}sign(\delta{x_{i}^{k}}(t))\frac{d
\delta{x_{i}^{k}}(t)}{dt}\nonumber\\
&
\le L\sum\limits_{i=1}^{m}
\theta_{i}||\delta{x}_{i}(t)||_{1}+c_{0}\sum\limits_{i,j=1}^{m}
\theta_{i}\gamma_{j}\tilde{a}_{ij}\sum\limits_{k=1}^{n}|\delta x_{j}^{k}(t)|-\sum\limits_{i=1}^{m}
\theta_{i}\nonumber\\
&
\le L\sum\limits_{i=1}^{m}
\theta_{i}||\delta{x}_{i}(t)||_{1}-c_{0}\theta\sum\limits_{j=1}^{m}
\gamma_{j}||\delta x_{j}(t)||_{1}-\sum\limits_{i=1}^{m}
\theta_{i}
\end{align*}

If $c_{0}\theta\min_{j}\{\gamma_{j}\}>L$, we have
\begin{align*}
\frac{d
||\delta{x}(t)||_{\{\theta,1\}}}{dt}< -\sum\limits_{i=1}^{m}\theta_{i}=-1
\end{align*}
which implies $||\delta{x}(t)||_{\{\theta,1\}}=0$, if $t\ge ||\delta{x}(0)||_{\{\theta,1\}}.$

In case $\Gamma$ is not a positive diagonal matrix, we have
\begin{theorem} Suppose that  QUAD-condition (6) is satisfied, $P\Gamma=BB^{T}$ is semi-positive definite, and is positive definite in $ker(P\Phi)^{\bot}$,

Then, the controlled
system
\begin{align}\label{pinf2}
\dot{x}_{i}(t)-\dot{s}(t)&=f(x_{i}(t))-f(s(t))\nonumber\\
&+c_{0}\sum\limits_{j=1}^{m}\tilde{a}_{ij}\Gamma (x_{j}(t)-s(t))-\frac{ x_{i}(t)-s(t)}{||x_{i}(t)-s(t)||_{\{{P,2}\}}}
\end{align}
where
\begin{align}
||x_{i}(t)-s(t)||_{\{{P,2}\}}=(x_{i}(t)-s(t))^{T}(t)P(x_{i}(t)-s(t))
\end{align}
is synchronized to $s(t)$ in finite time if $c_{0}$ is sufficiently large.
\end{theorem}

{\bf Proof}~
In this case, define a norm in Rimannian metric as
$$||\delta x(t)||_{\{\xi,P,2\}}=(V(\delta{x}(t)))^{\frac{1}{2}},$$
where
\begin{eqnarray*}
V(\delta{x}(t))=\frac{1}{2}\sum_{i=1}^{m}\xi_{i}
\delta{x}_{i}^{T}(t)P\delta{x}_{i}(t)
\end{eqnarray*}

Differentiating $V(\delta{x}(t))$, by $f\in QUAD(P,\Phi,\eta)$, we have
\begin{align*}
\frac{d
V(\delta{x}(t))}{dt}&=\sum\limits_{i=1}^{m}\xi_{i}\delta{x}_{i}^{T}(t)P\frac{d
\delta x_{i}(t) }{dt}\\
&\le-\eta \sum\limits_{i=1}^{m}\xi_{i}\delta{x}_{i}^{T}(t)\delta
x_{i}(t)\\
&+\frac{1}{2}\sum\limits_{i=1}^{m}\xi_{i}\delta{x}_{i}^{T}(t)
[P\Phi+\Phi^{T} P^{T}]\delta
x_{i}(t)\\
& +c_{0}\sum\limits_{i,j=1}^{m}\delta{x}_{i}^{T}(t)
\xi_{i}\tilde{a}_{ij}P\Gamma\delta x_{j}(t)\\
&-\sum\limits_{i=1}^{m}\xi_{i}||\delta x_{i}(t)||_{\{{P,2}\}}
\end{align*}

It is easy to check that
\begin{enumerate}

\item
Because $P\Gamma=BB^{T}$ is semi-positive definite, and is positive definite in $ker(P\Phi)^{\bot}$. Therefore, there exists a constant $c>0$, such that
\begin{align}
&\frac{1}{2}\sum\limits_{i=1}^{m}\xi_{i}\delta{x}_{i}^{T}(t)
[P\Phi+\Phi^{T} P^{T}]\delta
x_{i}(t)\le c\sum\limits_{i=1}^{m}\xi_{i}\delta{x}_{i}^{T}(t)
P\Gamma\delta
x_{i}(t)\nonumber\\
&= c\sum\limits_{i=1}^{m}\xi_{i}\delta{y}_{i}^{T}(t)
\delta y_{i}(t)
\end{align}
where $\delta y_{i}(t)=B^{T}\delta x_{i}(t)$.
\item
\begin{align}
&c_{0}\sum\limits_{i,j=1}^{m}\delta{x}_{i}^{T}(t)
\xi_{i}\tilde{a}_{ij}P\Gamma\delta x_{j}(t)=c_{0}\sum\limits_{i,j=1}^{m}\delta{y}_{i}^{T}(t)
\xi_{i}\tilde{a}_{ij}\delta y_{j}(t)\nonumber\\
&=\frac{c_{0}}{2}\sum\limits_{i,j=1}^{m}\delta{y}_{i}^{T}(t)
[\xi_{i}\tilde{a}_{ij}+\tilde{a}_{ji}\xi_{j}]\delta y_{j}(t)\nonumber\\
&\le\frac{\mu{c_{0}}}{2}\sum\limits_{i,j=1}^{m}\delta{y}_{i}^{T}(t)
\delta y_{j}(t)
\end{align}

\end{enumerate}

If $c+\frac{\mu{c_{0}}}{2}\le 0$, we have
\begin{align*}
&\frac{d
V(\delta{x}(t))}{dt}
\le-\sum\limits_{i=1}^{m}\xi_{i}||\delta x_{i}(t)||_{2}
\end{align*}

Thus,
$$\frac{d}{dt}||\delta x(t)||_{\{\xi,P,2\}}=\frac{1}{2}V^{-1/2}(\delta x(t))
\frac{dV(\delta{x})}{dt}<-\frac{1}{2} c_{4}$$
and 
$||\delta x(t)||_{\{\xi,P,2\}}=0$ when
$$t\ge 2\frac{V^{1/2}(\delta(x(0))}{ c_{4}}$$

Theorem is proved completely.
\begin{remark}
The idea proposed here comes from [9], in which  by using the theory of Filippov, it is revealed that if the equilibrium $x^{*}$ lies in the discontinuity of the activation functions, for example, $f(x)=sign(x)$, then finite time convergence can be ensured (see [9], Th. 8).

The basic idea used in [9] is to find a distance function $dis(x(t),x^{*})$, and prove
$$\frac{d}{dt}dis(x(t),x^{*})\le -c$$
for some positive constant $c$.

Proposition 1. Suppose a non-negative function $V(t)$ satisfies
$$\frac{dV(x(t))}{dt}\le -c\mu(V(x(t)))$$
where $c>0$, function $\mu(V(t))>0$, $V(t)>0$; $\mu(0)=0$.

In case that
$$
\int_{0}^{a}\mu^{-1}(W)dW<\infty$$
define a distance function
$$
dis(V(x(t)),0)=\int_{0}^{V(x(t))}\mu^{-1}(W)dW$$
It is easy to check that
$$\frac{d
\int_{0}^{V(x(t))}\mu^{-1}(W)dW}{dt}
=\mu^{-1}(V(x(t)))\frac{dV(x(t))}{dt}\le -c$$
which implies $V(x(t))$ converges to zero in finite time
$\tilde{t}=\frac{\int_{0}^{V(0)}\mu(W)^{-1}dW}{c}$.

\end{remark}

\section{Distributed Finite Time Synchronization algorithms of Complex Networks with single pinning linear and nonlinear controller}

In this section, we introduce distributed discontinuous and continuous algorithms for finite time synchronization of complex networks with single pinning linear and nonlinear controller.

By a coordinate transform  $y_{i}(t)=\sum\limits_{j=1}^{m}\tilde{a}_{ij} x_{j}(t)$, $\delta{y}_{i}(t)=\sum\limits_{j=1}^{m}\tilde{a}_{ij} (x_{j}(t)-s(t))$. Because $(\tilde{a}_{ij})$ is a nonsingular M-matrix, $\delta{y}_{i}(t)\rightarrow 0$ is equivalent to $\delta{x}_{i}(t)\rightarrow 0$.

Replacing $\delta{x}_{i}(t)$ by $\delta{y}_{i}(t)$, we propose following distributed algorithm
\begin{eqnarray}\label{pinf}
\left\{\begin{array}{ll}\frac{dx_1(t)}{dt}&=f(x_1(t),t)
+c_{01}\sum\limits_{j=1}^m\tilde{a}_{1j} (x_j(t)-s(t))\\
&+\frac{\sum\limits_{j=1}^m\tilde{a}_{1j}(x_j(t)-s(t))}
{||\sum\limits_{j=1}^m\tilde{a}_{1j}(x_j(t)-s(t))||_{1}},\\
\frac{dx_i(t)}{dt}&=f(x_i(t),t)+c_{01}\sum\limits_{j=1}^ma_{ij} x_j(t)+\frac{\sum\limits_{j=1}^ma_{ij}x_j(t)}
{||\sum\limits_{j=1}^ma_{ij}x_j(t)||_{1}},\\
&i=2,\cdots,m \end{array}\right.
\end{eqnarray}
\begin{theorem} Suppose that $\Gamma=diag[\gamma_{1},\cdots,\gamma_{m}]$ is a positive diagonal matrix. $||f(x)-f(y)||_{1}\le L||x-y||_{1}$.
Then, the controlled system (12)
can synchronize to $s(t)$ in finite time if $c_{0}\theta>L$. Here, $||x_{i}(t)-s(t)||_{1}=\sum_{k=1}^{n}|x_{i}^{k}(t)-s^{k}(t)|$. 
\end{theorem}
{\bf Proof}~
In this case, system (12) can be written as
\begin{align}
\dot{\delta}y_{i}(t)=&\sum\limits_{j=1}^{m}\tilde{a}_{ij}\delta f(x_{j}(t))
+c_{01}\sum\limits_{j=1}^{m}\tilde{a}_{ij} \delta{y}_{j}(t)\nonumber\\
&+\sum\limits_{j=1}^{m}\tilde{a}_{ij} \frac{\delta{y}_{j}(t)}{||\delta{y}_{j}(t)||_{1}}
\end{align}

Similar to the proof of Theorem 1, define a norm
\begin{eqnarray*}
||\delta{y}(t)||_{\{\theta,1\}}=\sum_{i=1}^{m}\theta_{i}
||\delta{y}_{i}(t)||_{1}=\sum_{i=1}^{m}\theta_{i}
\sum_{k=1}^{n}|\delta{y}_{i}^{k}(t)|
\end{eqnarray*}
where $\delta{y}_{i}(t)=[\delta{y}_{i}^{1}(t),\cdots,\delta{y}_{i}^{n}(t)]^T$,
$i=1,\cdots,m$, $\delta{y}(t)=[\delta{y}_{1}^T(t),\cdots,\delta{y}_{m}^T(t)]^{T}$.

It is easy to see that
\begin{align*}
\sum_{i=1}^{m}\theta_{i}|\sum\limits_{j=1}^{m}\tilde{a}_{ij}\delta f(x_{j}(t))|&\le L\sum_{i=1}^{m}\theta_{i}\sum\limits_{j=1}^{m}|\tilde{a}_{ij}|||\delta x_{j}(t)||_{1}
\\&\le \tilde{L}\sum_{i=1}^{m}\theta_{i}
||\delta y_{i}(t)||_{1}
\end{align*}
and
\begin{align*}
\frac{d
||\delta{y}(t)||_{\{\theta,1\}}}{dt}&\le \tilde{L}\sum_{i=1}^{m}\theta_{i}
||\delta y_{i}(t)||_{1}+c_{01}\sum\limits_{i,j=1}^{m}
\theta_{i}\tilde{a}_{ij}||\delta y_{j}(t)||_{1}\nonumber\\
&+\sum\limits_{j=1}^{m}
\sum\limits_{i=1}^{m}\theta_{i}\tilde{a}_{ij}
\end{align*}
In case that $c_{01}\theta>\tilde{L}\max_i\theta_i$, we have
\begin{align*}
\frac{d
||\delta{y}(t)||_{\{\theta,1\}}}{dt}&<-m\theta <0
\end{align*}

Therefore,
$$||\delta{y}(t)||_{\{\theta,1\}}=0$$
when $t\ge \frac{||\delta{y}(0)||_{1}}{m\theta }$.

Now, we propose following continuous finite time convergence algorithm
\begin{eqnarray}
\left\{\begin{array}{ll}\frac{dx_1(t)}{dt}&=f(x_1(t),t)
+c_{01}\sum\limits_{j=1}^m\tilde{a}_{1j} (x_j(t)-s(t))\\
&+\frac{\sum\limits_{j=1}^m\tilde{a}_{1j}(x_j(t)-s(t))}
{||\sum\limits_{j=1}^m\tilde{a}_{1j}(x_j(t)-s(t))||_{1}^{\alpha}},\\
\frac{dx_i(t)}{dt}&=f(x_i(t),t)+c_{01}\sum\limits_{j=1}^ma_{ij} x_j(t)+\frac{\sum\limits_{j=1}^ma_{ij}x_j(t)}
{||\sum\limits_{j=1}^ma_{ij}x_j(t)||_{1}^{\alpha}},\\
&i=2,\cdots,m \end{array}\right.
\end{eqnarray}
where $0<\alpha<1$.

By similar derivations, we have
\begin{align*}
\frac{d
||\delta{y}(t)||_{\{\theta,1\}}}{dt}&\le \tilde{L}\sum_{i=1}^{m}\theta_{i}
||\delta y_{i}(t)||_{1}+c_{01}\sum\limits_{i,j=1}^{m}
\theta_{i}\tilde{a}_{ij}||\delta y_{j}(t)||_{1}\nonumber\\
&+\sum\limits_{j=1}^{m}
\sum\limits_{i=1}^{m}\theta_{i}\tilde{a}_{ij}||\delta{y}_{j}(t)||_{1}^{1-\alpha}
\end{align*}
In case that $c_{01}\theta>\tilde{L}\max_i\theta_i$, we have
\begin{align*}
\frac{d||\delta{y}(t)||_{\{\theta,1\}}}{dt}&<-\theta \sum\limits_{j=1}^{m}||\delta{y}_{j}(t)||_{1}^{1-\alpha}
\end{align*}

By Jensen inequality
\begin{align*}
(\sum\limits_{j=1}^{m}||\delta{y}_{j}(t)||_{1})^{1-\alpha}\le m^\alpha(\sum\limits_{j=1}^{m}||\delta{y}_{j}(t)||_{1}^{1-\alpha})
\end{align*}
we have
\begin{align*}
\frac{d
||\delta{y}(t)||_{\{\theta,1\}}}{dt}&<-\frac{\theta }{m^\alpha}(\sum\limits_{j=1}^{m}||\delta{y}_{j}(t)||_{1})^{1-\alpha}
\le -\frac{\theta }{m^\alpha}||\delta{y}(t)||_{\{\theta,1\}}^{1-\alpha}
\end{align*}
and by Proposition 1, $\delta y(t)$ converges to zero in finite time.

\section{Heterogeneous dynamic networks}

As another application of previous results, in this section, we discuss synchronization for heterogeneous dynamic networks.

Given $m+1$ systems $\dot{x}(t)=f_{i}(x(t))$, $i=0,1,\cdots,m$, where $f_{i}(x)$, $i=0,1,\cdots,m$, are (maybe different) bounded continuous functions.

Consider following system
\begin{eqnarray}\label{pinfl3b}
\left\{\begin{array}{ll}\frac{dx_1(t)}{dt}&=f(x_1(t),t)
+c_{01}\sum\limits_{j=1}^m\tilde{a}_{1j} (x_j(t)-s(t))\\
&+c_{02}\frac{\sum\limits_{j=1}^m\tilde{a}_{1j}(x_j(t)-s(t))}
{||\sum\limits_{j=1}^m\tilde{a}_{1j}(x_j(t)-s(t))||_{1}},\\
\frac{dx_i(t)}{dt}&=f(x_i(t),t)+c_{01}\sum\limits_{j=1}^ma_{ij} x_j(t)+c_{02}\frac{\sum\limits_{j=1}^ma_{ij}x_j(t)}
{||\sum\limits_{j=1}^ma_{ij}x_j(t)||_{1}},\\
&i=2,\cdots,m \end{array}\right.
\end{eqnarray}
where $s(t)$ is a solution satisfying $\dot{s}(t)=f_{0}(s(t))$.
\begin{theorem}
Algorithm (18) can reach synchronization in finite time for heterogeneous networks.
\end{theorem}
{\bf Proof} Algorithm (18) can be rewritten as
\begin{align}\label{pinfl3c}
\dot{\delta}y_{i}(t)=&\sum\limits_{j=1}^{m}\tilde{a}_{ij}\delta f(x_{j}(t))
+c_{01}\sum\limits_{j=1}^{m}\tilde{a}_{ij} \delta{y}_{j}(t)\nonumber\\
&+c_{02}\sum\limits_{j=1}^{m}\tilde{a}_{ij} \frac{\delta{y}_{j}(t)}{||\delta{y}_{j}(t)||_{1}}+u_{i}(t)
\end{align}
where $u_{i}(t)=f_{i}(x_i(t),t)-f_{0}(s(t))$ and satisfies $||u_{i}(t)||\le c$.

Similar to the proof of Theorem 3, define
\begin{align}
V(x(t))=\sum_{i=1}^{m}\theta_{i}||x_i(t)-s(t)||_{1}
\end{align}
In case that $c_{01}\theta>\tilde{L}\max_i\theta_i$, aand $c_{02}>\frac{c}{m\theta}$, we have
\begin{align*}
\frac{d
||\delta{y}(t)||_{\{\theta,1\}}}{dt}&<-m\theta c_{02}+c<0
\end{align*}
Therefore, $||\delta{y}(t)||_{\{\theta,1\}}=0$, when $t>\frac{||\delta{y}(0)||_{\{\theta,1\}}}{m\theta c_{02}-c}$.

\section{Finite Time Pinning Consensus of Complex Networks}
In case that $f(x(t),t)=\bar{A}x(t)$, $\dot{s}(t)=\bar{A}s(t)$, it is clear there exists a constant $L_{\bar{A}}$ such that $||\bar{A}(x-y)||_{1}\le L_{\bar{A}}||x-y||_{1}$.

As a direct consequence of Theorem 1, we can give
\begin{theorem}
The controlled system
\begin{eqnarray}\label{pinfl3b}
\left\{\begin{array}{ll}\frac{dx_1(t)}{dt}&=\bar{A}x_1(t)
+c_{01}\sum\limits_{j=1}^m\tilde{a}_{1j} (x_j(t)-s(t))\\
&+\frac{\sum\limits_{j=1}^m\tilde{a}_{1j}(x_j(t)-s(t))}
{||\sum\limits_{j=1}^m\tilde{a}_{1j}(x_j(t)-s(t))||_{1}},\\
\frac{dx_i(t)}{dt}&=\bar{A}x_i(t)+c_{01}\sum\limits_{j=1}^ma_{ij} x_j(t)+\frac{\sum\limits_{j=1}^ma_{ij}x_j(t)}
{||\sum\limits_{j=1}^ma_{ij}x_j(t)||_{1}},\\
&i=2,\cdots,m \end{array}\right.
\end{eqnarray}
can synchronize all $x_{i}(t)$ to $s(t)$ in finite time.
\end{theorem}


\section{Simulations}
In this section, we give several simulations to verify previous theoretical analysis.

We assume that $s(t)$ is the Lorenz system described by $\dot{s}(t)=f(s(t))$, where
\begin{align}
f(s(t))=\left(
\begin{array}{c}
10(s^2(t)-s^1(t))\\
28s^1(t)-s^2(t)-s^1(t)s^3(t)\\
s^1(t)s^2(t)-8s^3(t)/3
\end{array}\right)
\end{align}
Coupling matrix is
\begin{align}
(a_{ij})_{i,j=1}^{m}=\left(
    \begin{array}{ccc}
    -3  &   1  &   2\\
     2  &  -4  &   2\\
     2  &   1  &  -3 \\
    \end{array}
  \right)
\end{align}
$\epsilon=1$, and
\begin{align*}
E(t)=\|(x_1(t)^T,x_2(t)^T,x_3(t)^T)^T-(1, 1, 1)^T\otimes s(t)\|_1
\end{align*}
is used to verify effectiveness of the algorithms.

{\bf Simulation 1.}
In this simulation, we consider nonlinear feedback controllers with a single linear feedback controller
\begin{align}\label{simu1}
\left\{
\begin{array}{ll}
\dot{x}_1(t)=&f(x_1(t))+5\sum_{j=1}^3\tilde{a}_{1j}x_j(t)
-5\frac{x_1(t)-s(t)}{\|x_1(t)-s(t)\|_1}\\
\dot{x}_i(t)=&f(x_i(t))+5\sum_{j=1}^3a_{ij}x_j(t)-5\frac{x_i(t)-s(t)}{\|x_i(t)-s(t)\|_1},\\
&~~~~~~~~~~~~~~~~~~~~~~~~~~~~~~~~~~~~~~~~~~~i=2,3
\end{array}
\right.
\end{align}
Dynamical behavior $E(x(t))$ is depicted in Figure 1.

\begin{figure}
\begin{center}
\includegraphics[width=0.5\textwidth]{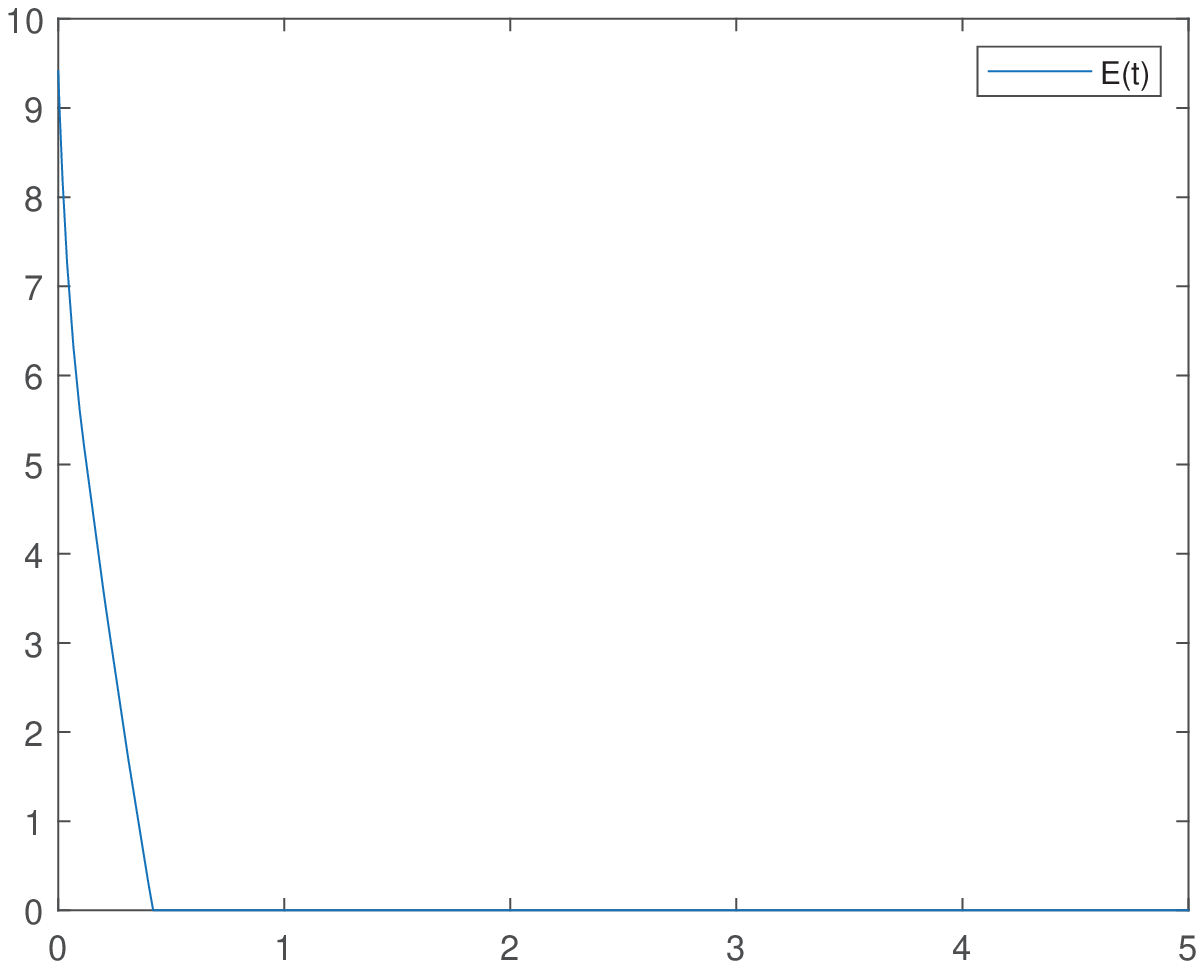}
\caption{Synchronization error $E(t)$ based on model (21)}
\end{center}
\end{figure}

{\bf Simulation 2.} Synchronization by a single pinning linear feedback controller with a single nonlinear feedback controller 
\begin{eqnarray}\label{simu2}
\left\{\begin{array}{ll}\frac{dx_1(t)}{dt}&=f(x_1(t),t)
+5\sum\limits_{j=1}^m\tilde{a}_{1j} (x_j(t)-s(t))\\
&+5\frac{\sum\limits_{j=1}^m\tilde{a}_{1j}(x_j(t)-s(t))}
{||\sum\limits_{j=1}^m\tilde{a}_{1j}(x_j(t)-s(t))||_{1}},\\
\frac{dx_i(t)}{dt}&=f(x_i(t),t)+5\sum\limits_{j=1}^ma_{ij} x_j(t)+5\frac{\sum\limits_{j=1}^ma_{ij}x_j(t)}
{||\sum\limits_{j=1}^ma_{ij}x_j(t)||_{1}},\\
&i=2,3 \end{array}\right.
\end{eqnarray}

Dynamical behavior $E(x(t))$ is depicted in Figure 2.
\begin{figure}
\begin{center}
\includegraphics[width=0.5\textwidth]{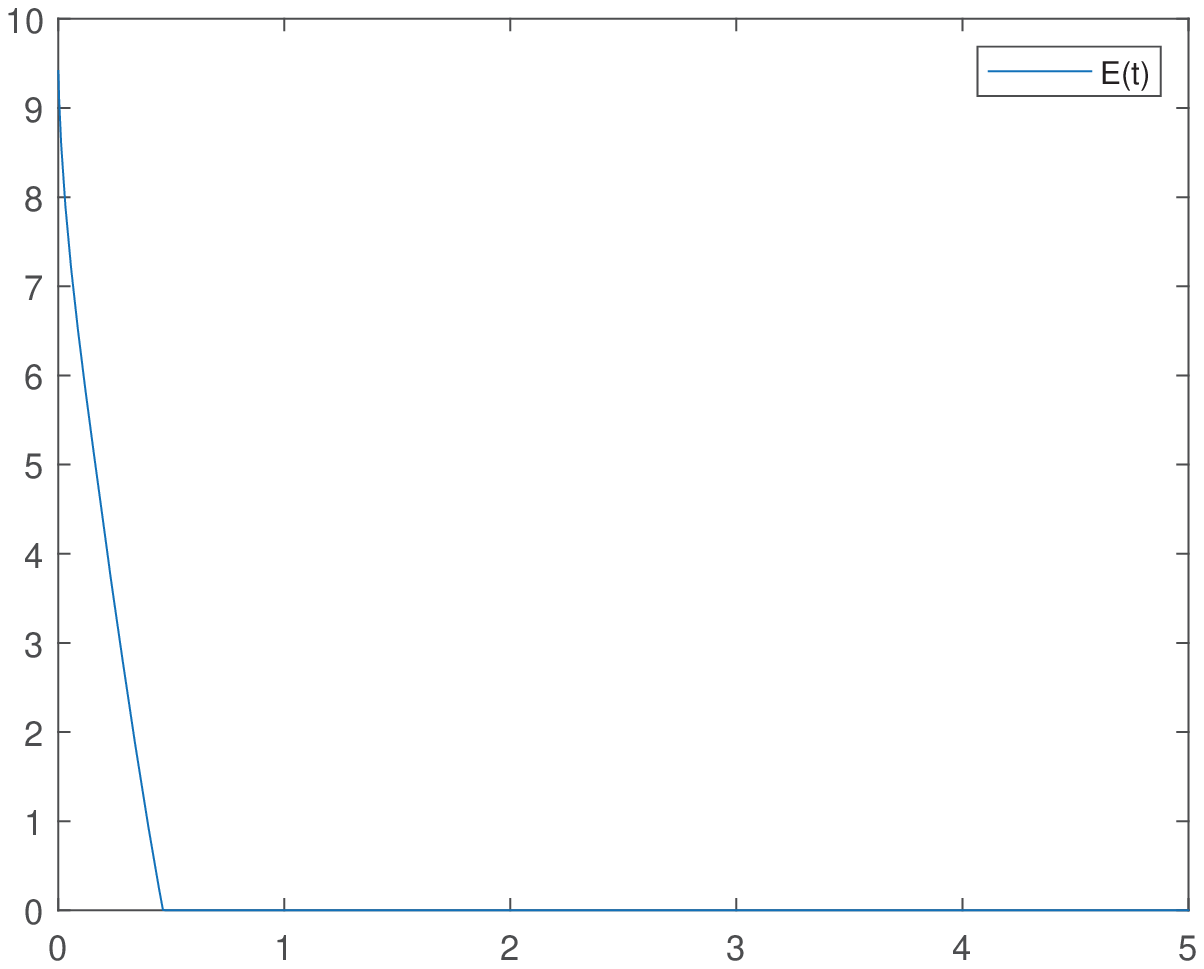}
\caption{Synchronization error $E(t)$ based on model (22)}
\end{center}
\end{figure}

{\bf Simulation 3.} Pinning synchronization model 
\begin{eqnarray}\label{simu3}
\left\{\begin{array}{ll}\frac{dx_1(t)}{dt}&=f(x_1(t),t)
+5\sum\limits_{j=1}^m\tilde{a}_{1j} (x_j(t)-s(t))\\
&+5\frac{\sum\limits_{j=1}^m\tilde{a}_{1j}(x_j(t)-s(t))}
{||\sum\limits_{j=1}^m\tilde{a}_{1j}(x_j(t)-s(t))||_{1}^{1/2}},\\
\frac{dx_i(t)}{dt}&=f(x_i(t),t)+5\sum\limits_{j=1}^ma_{ij} x_j(t)+5\frac{\sum\limits_{j=1}^ma_{ij}x_j(t)}
{||\sum\limits_{j=1}^ma_{ij}x_j(t)||_{1}^{1/2}},\\
&i=2,3 \end{array}\right.
\end{eqnarray}

Dynamical behavior $E(x(t))$ is depicted in Figure 3.
\begin{figure}
\begin{center}
\includegraphics[width=0.5\textwidth]{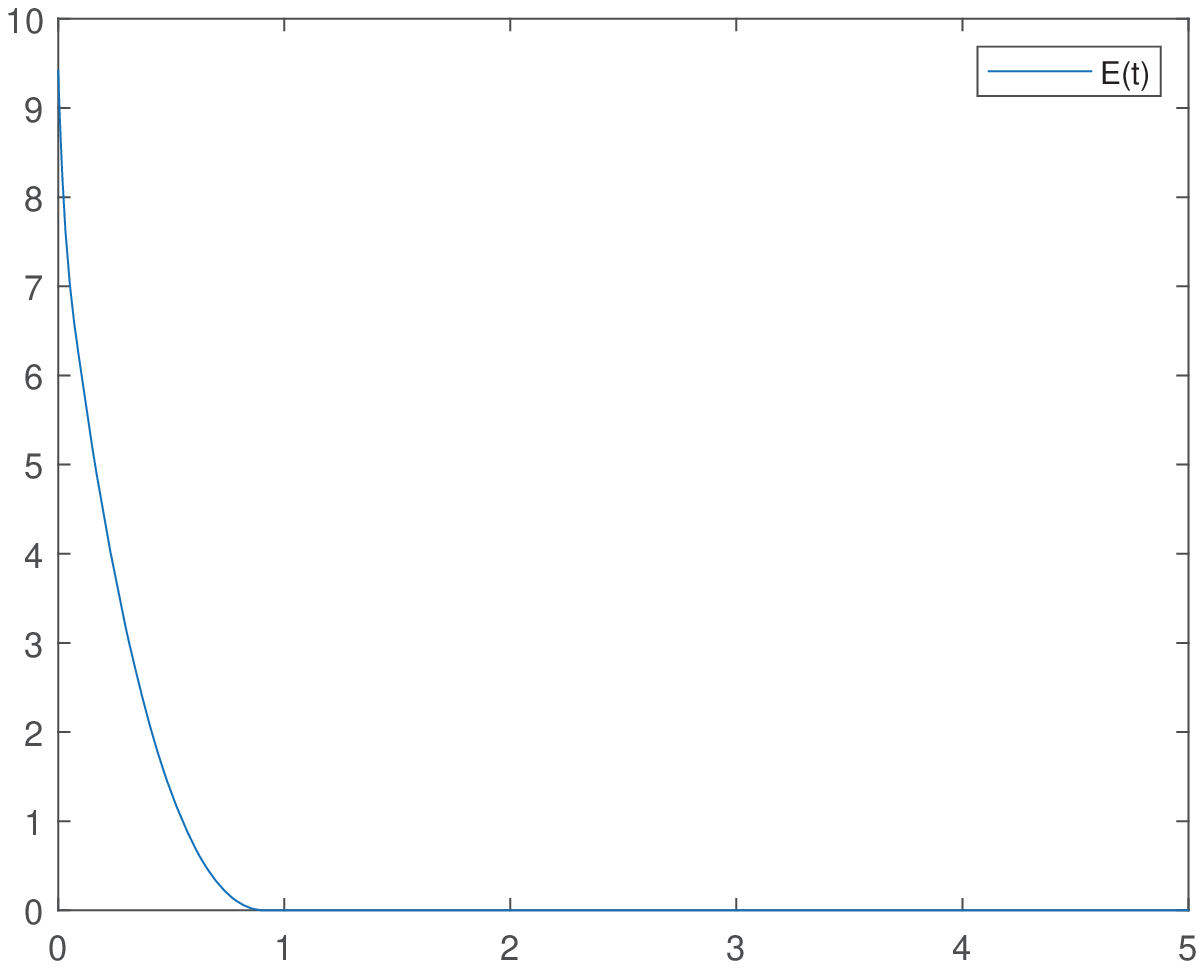}
\caption{Synchronization error $E(t)$ based on model (23)}
\end{center}
\end{figure}

{\bf Simulation 4.} Synchronization for heterogeneous systems
\begin{eqnarray}\label{heteo}
\left\{\begin{array}{ll}\frac{dx_1(t)}{dt}&=f_{1}(x_1(t),t)
+5\sum\limits_{j=1}^m\tilde{a}_{1j} (x_j(t)-s(t))\\
&+5\frac{\sum\limits_{j=1}^m\tilde{a}_{1j}(x_j(t)-s(t))}
{||\sum\limits_{j=1}^m\tilde{a}_{1j}(x_j(t)-s(t))||_{1}},\\
\frac{dx_i(t)}{dt}&=f_{i}(x_i(t),t)+5\sum\limits_{j=1}^ma_{ij} x_j(t)+5\frac{\sum\limits_{j=1}^ma_{ij}x_j(t)}
{||\sum\limits_{j=1}^ma_{ij}x_j(t)||_{1}},\\
&i=2,3 \end{array}\right.
\end{eqnarray}
where $s(t)$ is the Lorenz system described by $\dot{s}(t)=f(s(t))$ satisfying
\begin{align}
f(s(t))=\left(
\begin{array}{c}
10(s^2(t)-s^1(t))\\
28s^1(t)-s^2(t)-s^1(t)s^3(t)\\
s^1(t)s^2(t)-8s^3(t)/3
\end{array}\right)
\end{align}
$f_{1}(x(t)$ is Chua circuit, $f_{2}(x(t))=f_{1}(x(t)+sin(t)$, $f_{3}(x(t))=f_{1}(x(t)+cos(t)$.

Dynamical behavior $E(x(t))$ is depicted in Figure 4.
\begin{figure}
\begin{center}
\includegraphics[width=0.5\textwidth]{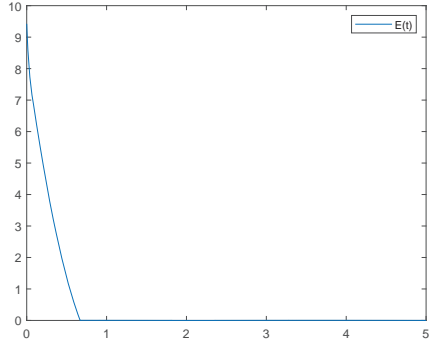}
\caption{Synchronization error $E(t)$ based on heterogeneous model (24)}
\end{center}
\end{figure}

\end{document}